\documentstyle[12pt,aasms4]{article}
\newcommand{\postscript}[2]
 {\setlength{\epsfxsize}{#2\hsize}
   \centerline{\epsfbox{#1}}}

\begin{document}

\title{\sc Chromaticity of Gravitational Microlensing Events}

\bigskip
\bigskip
%---authors
\author{Cheongho Han}
\smallskip
\author{Seong-Hong Park}
\smallskip
\centerline{\&}
\author{Jang-Hae Jeong}
\bigskip
\affil{Department of Astronomy \& Space Science, \\
       Chungbuk National University, Chongju, Korea 361-763 \\
       cheongho,parksh@astronomy.chungbuk.ac.kr\\
       jeongjh@astro.chungbuk.ac.kr}
\authoremail{cheongho,parksh@astronomy.chungbuk.ac.kr}
\authoremail{jeongjh@astro.chungbuk.ac.kr}

\bigskip
\bigskip

%======= ABSTRACT =================================================
\begin{abstract}
In this paper, we investigate the color changes of gravitational microlensing 
events caused by the two different mechanisms of differential amplification for 
a limb-darkened extended source and blending. From this investigation, we find 
that the color changes of limb-darkened extended source events (color curves) 
have dramatically different characteristics depending on whether the lens 
transits the source star or not. We show that for a source transit event, 
the lens proper motion can be determined by simply measuring the turning time 
of the color curve instead of fitting the overall color or light curves. We 
also find that even for a very small fraction of blended light, the color 
changes induced by the blending effect is equivalent to those caused by the 
limb-darkening  effect, causing serious distortion in the observed color curve. 
Therefore, to obtain useful information about the lens and source star from 
the color curve of a limb-darkened extended source event, it will be essential 
to eliminate or correct for the blending effect. We discuss about the methods 
for the efficient correction of the blending effect.
\end{abstract}

\bigskip
\bigskip
\keywords{gravitational lensing -- stars: giants, main sequence -- 
limb darkening}

\centerline{resubmitted to {\it Monthly Notices of the Royal Astronomical
Society}: Nov 20, 1999}
\centerline{Preprint: CNU-A\&SS-10/99}
\clearpage

%==== Main Body ==================================================
\section{Introduction}

Searches for Galactic dark matter by monitoring light variations of stars 
caused by gravitational microlensing have been and are being carried out 
towards the Galactic bulge and the Magellanic Cloud fields by several groups 
(MACHO: Alcock et al.\ 1993; OGLE: Udalksi et al.\ 1993; EROS: Aubourg et al.\ 
1993; DUO: Alard \& Guibert 1997).  With their efforts, $\sim 500$ events have 
been detected to date.

One of the important characteristics of gravitational microlensing 
amplification is that the color of the light curve does not change during the 
event.  However, the achromaticity of the light curve is violated for some 
cases of gravitational microlensing events. The first mechanism that makes 
the lensing event light curve wavelength dependent is the differential 
amplification for a limb-darkened extended source (see \S\ 2).  By measuring 
the color changes induced by this mechanism (hereafter color curve), one can 
not only investigate the intensity profile of the source star but also 
determine the lens proper motion from which the mass and location of the 
lens can be significantly better determined.\footnote{Investigation of the 
intensity profile of the source star and determination of the lens proper 
motion are also possible by analyzing a single band {\it light curve} of the 
event. This is because not only the color but also the amplification is 
affected by the extended source effect.  However, if the lens does not transit 
the source star, the amplification change induced by the extended source effect
simply masquerades as changes in lensing parameters, and thus it cannot be 
detected. By contrast, the color effects cannot be mimicked by the changes in 
lensing parameters, since point-source lensing event is achromatic. Gould 
\& Welch (1996) showed that by detecting the color changes, one can double 
the chance to detect the extended source effect.}  Another mechanism making 
the lensing light curve chromatic is blending.  If the measured flux of the 
source star is affected by the blended light from unresolved stars with 
different colors, the color of the light curve is changed during the event 
(see \S\ 3).

In this paper, we investigate the color changes of gravitational microlensing 
events caused by the two different mechanisms of differential amplification 
for a limb-darkened extended source and blending.  From this investigation, we 
find that the color changes of limb-darkened extended source events  
have dramatically different characteristics depending on whether the lens 
transits the source star or not.  We show that for a source transit event, 
the lens proper motion can be determined by simply measuring the turning time 
of the color curve instead of fitting the overall color or light curves.  We 
also find that even for a very small fraction of blended light, the color 
changes induced by the blending effect is equivalent to those caused by the 
limb-darkening  effect, causing serious distortion in the observed color curve. 
Therefore, to obtain useful information about the lens and source star from 
the color curve of a limb-darkened extended source event, it will be essential 
to eliminate or correct for the blending effect. We discuss about the methods 
for the efficient correction of the blending effect.

\section{Color Change Caused by Limb Darkening}

The light curve of an event with an isolated point source is represented by
$$
A_0 = {u^2+2\over u(u^2+4)^{1/2}},
\eqno(2.1)
$$
where $u$ is the lens-source separation in units of the angular Einstein ring 
radius $\theta_{\rm E}$. The Einstein ring represents the effective region of 
gravitational lensing amplification within which the source star amplification 
becomes greater than $\sqrt{5}/3$. The angular Einstein ring radius is related 
to the lens mass $M$ and its location by
$$
\theta_{\rm E}=\left({4GM\over c^2}{D_{ls}\over D_{ol}D_{os}}\right)^{1/2},
\eqno(2.2)
$$
where $D_{ol}$, $D_{ls}$, and $D_{os}$ are the separations between the 
observer, lens, and source star, respectively. The value of $\theta_{\rm E}$ 
for a Galactic bulge event caused by a solar mass lens located at the 
half way between the source star and observer (i.e.\ $D_{ol}/D_{os}=0.5$) 
is $\theta_{\rm E}\sim 0.3$ milli-arcsec. On the other hand, typical 
main-sequence stars in the Galactic bulge have radii that extend only 
$\theta_\ast\lesssim 1$ $\mu$-arcsec. Therefore, equation (2.1) is a good
approximation for most of Galactic microlensing events. The gravitational 
amplification for a point source event is characterized by a symmetric, 
non-repeating, and achromatic light curve.

However, for a very close lens-source impact event with a considerable source 
star size such as giants, the source star can no longer be approximated by 
a point source. For an extended source event, the light curve is given by 
the intensity-weighted amplification averaged over the surface of the source 
star, i.e.\ 
$$
A_\nu = {
\int_0^{2\pi}\int_0^{r_\ast} I_\nu(r,\vartheta) A_0(\left\vert 
{\bf r} - {\bf r}_{L}\right\vert) r dr d\vartheta
\over 
\int_0^{2\pi}\int_0^{r_\ast} I_\nu(r,\vartheta)r dr d\vartheta
},
\eqno(2.3)
$$
where $r_\ast$ is the radius of the source star, $I_{\nu}(r,\vartheta)$ is 
the surface intensity distribution of the source star, and the vectors 
${\bf r}_{L}$ and ${\bf r}=(r,\vartheta)$ represent the displacement vector 
of the center of source star with respect to the lens and the orientation 
vector of a point on the source star surface with respect to the center of 
the source star, respectively.  For the radially symmetric distribution of 
the source star intensity, equation (2.3) is simplied into
$$
A_\nu = {
\int_{0}^{r_\ast} I_\nu(r)A_0(\left\vert {\bf r}-{\bf r}_{L}\right\vert)rdr
\over 
\int_{0}^{r_\ast} I_\nu(r)rdr
}
.
\eqno(2.4)
$$

The light curve of an extended source event can become chromatic due to limb 
darkening. Because observations at different wavelengths probe material at 
different depth in stellar atmosphere, the radial surface intensity profile 
of a star is wavelength dependent. In longer wavelength bands, which probe 
the cooler outer regions of the star, the stellar disk will appear less limb 
darkened. When the lens passes close to the source star, different parts of 
the source star disk (with varying surface intensity and spectral energy 
distribution) are amplified by different amount due to the differences in 
distance to the lens: differential amplification (Gould 1994; Nemiroff \& 
Wickramasinghe 1994; Witt \& Mao 1994). As a result, the microlensing event 
with a limb-darkened extended source star will have different amplifications 
depending on the observed wavelength bands. Let us define $F_{\nu i}=2\pi
\int_{0}^{r_\ast} I_{\nu i}(r)rdr;$ $i$=1, 2 and $m_{\nu i}$ as the unamplified
source star fluxes and the corresponding maginitudes measured in two different 
wavelength bands $\nu 1$ and $\nu 2$. Then the color changes during the lensing
event caused by the limb darkening of the source star are computed by
$$
\Delta(m_{\nu 2}-m_{\nu 1}) = 
(m_{\nu 2}-m_{\nu 1}) - (m_{\nu 2}-m_{\nu 1})_0 
= -2.5 \log \left( {A_{\nu 2} \over A_{\nu 1}}\right),
\eqno(2.5)
$$
where $(m_{\nu 2}-m_{\nu 1})_0 = -2.5 \log (F_{\nu 2}/F_{\nu 1})$ and 
$m_{\nu 2}-m_{\nu 1} = -2.5 \log (A_{\nu 2}F_{\nu 2}/A_{\nu 1}F_{\nu 1})$ 
represent the color differences before and during the gravitational 
amplification. From equation (2.5), one finds that the color change caused 
by the limb darkening effect depends on the ratio of the amplifications 
observed in two bands $A_{\nu 2}/A_{\nu 1}$, but does not depends on the 
source star fluxes $F_{\nu i}$.

\subsection{Types of Color Curve Patterns}

To see the pattern of the color curves, we compute the color changes of 
example events for a limb-darkened source star with an angular radius of 
$\theta_\ast=0.1\theta_{\rm E}$ and the resulting light curves are presented 
in Figure 1. On the left side, the trajectories of the lens with respect to 
the source star (the shaded circle) of the individual events are marked by 
straight lines with different line types. In the right panel, the resulting 
color curves $\Delta(U-I)$ for the individual events are drawn by the same 
line types as those of the corresponding trajectories in the left panel. The 
unamplified color of the source star is $(U-I)_0=2.98$, which corresponds to 
that of a K-type giant (Allen 1973; Schmidt-Kaler 1982; Peletier 1989). For 
the surface intensity profile, we adopt a linear form of
$$
I_\nu (r) = 1- {\cal C}_\nu \left[ 1- \sqrt{1-(r/r_\ast)^2}\right],
\eqno(2.1.1)
$$
where the limb-darkening coefficients are ${\cal C}_\nu = 1.050$ and 0.503 in 
$U$ and $I$ bands, respectively, which correspond to those of a K giant with 
$T_{\rm eff}=4,750$ K, $\log g \sim 2.0$, and a metallicity similar to the 
sun (Van Hamme 1993).

From Figure 1, one finds that the patterns of the color curve can be 
classied broadly into two categories depending on whether the lens transits 
the source star disk or not. First, if the lens transits the source star 
(transit event), the resulting color curve is characterized by the turn of 
the color curve during the event. As the lens approaches the source star 
but before the transit, the color of the source star becomes redder because 
the closer (and thus more amplified) part of the source star to the lens is 
the cooler outer part. The source star appears to be most reddish at the 
moment when the lens is located at the limb of the star, and thus the effect 
of differential amplification is maximized. As the lens further approaches and 
transits the source star, on the other hand, the closer part of the source 
star to the lens is the hotter inner region, causing a turn in the color 
curve. Second, if the lens passes close to the limb of the source star but 
does not actually transit (passing event), the color of the source star 
continues to become redder without any turn as the lens approaches the source 
star. In addition, the amount of the color changes for passing events are 
smaller than those of crossing events.

\subsection{Source-Transit Event Rate}

An interesting finding from the color curve of a transit event is that one
can determine the lens proper motion by simply measuring the turning time
$t_\cap$ instead of fitting the overall color or light curves. The turn of the 
color change occurs when $t_\cap=\pm\left[ (\theta_\ast/\theta_{\rm E})^2- 
\beta^2 \right]^{1/2} t_{\rm E}$. Since the values of the lensing parameters 
$\beta$ and $t_{\rm E}$ can be measured from the overall shape of the light 
curve, one can determine the angular Einstein ring radius by
$$
\theta_{\rm E}={\theta_\ast \over
[\beta^2+(\left\vert t_\cap\right\vert/t_{\rm E})^2]^{1/2}}.
\eqno(2.2.1)
$$
Determining $\theta_{\rm E}$ is equivalent to the determination of the lens 
proper motion $\mu$ relative to the observer-source line of sight because 
$\mu=\theta_{\rm E} /t_{\rm E}$.

Then what will be the fraction of transit events out of total Galactic 
bulge events with giant source stars?  Another question that should be 
answered is for what fraction of these source-transit events can be actually 
detected by the current lensing experiments.  To answer these questions, we 
compute the event rate as a function of the Einsten time scale for Galactic 
bulge events with giant source stars by 
$$
\Gamma(t_{\rm E}) = 
\int_0^\infty dD_{\rm os}\rho(D_{os})
\int_0^{D_{os}} dD_{\rm ol}\rho(D_{ol})
\int_0^\infty \int_0^\infty  dv_y dv_z v f(v_y,v_z)
$$
$$
\times
\int_0^\infty dM \left( {4GM\over c^2} {D_{ol}D_{ls}\over D_{os}}\right)^{1/2}
\Phi(M) \delta \left[ t_{\rm E}-\left( {4GM \over c^2 v^2} {D_{ol}D_{ls}
\over D_{os}} \right)^{1/2}\right],
\eqno(2.2.2)
$$
where $\rho(D_{os})$ and $\rho(D_{ol})$ are the number densities of giant 
source stars and lenses along the line of sight towards the Galactic bulge, 
$(v_y,v_z)$ are the two components of the lens-source transverse velocity 
{\bf v}, $f(v_y,v_z)$ represents their distribution, and $\Phi(M)$ is the 
mass function of lenses.  For the matter distribution, we adopt a `revised 
COBE' Galactic bulge model (Dwek et al.\ 1995) and a double-exponential 
Galactic disk model (Bahcall 1986).  The transverse velocity distribution 
is modeled by a Gaussian.  The detailed descriptions about the matter and 
transverse velocity distributions are found in Han \& Gould (1996).  The 
lens mass function is modeled by a power law with a mass cutoff, i.e.\ 
$\Phi(M) \propto M^{-p} \Theta (m-m_{\rm cut})$, where the adopted values of 
the power and the cutoff mass are respectively $p=2.1$ and $m_{\rm cut}=0.04\ 
M_\odot$, following the determination of Han \& Gould (1996).  The relative 
event rate distribution for the total Galactic bulge events with giant source 
stars is presented in Figure 2 (dashed curve).

To become a transit event, the source star radius normalized by 
$\theta_{\rm E}$ should be greater than the lens-source impact parameter of 
the event, i.e.\
$$
\beta_{\rm crit} > \beta; \qquad 
\beta_{\rm crit}={\theta_\ast\over\theta_{\rm E}}
\eqno(2.2.3)
$$
(see Figure 3).  With this definition of a source-transit event and the 
adopted source star radius of a K-giant, we determine the event rate 
distribution for source-transit events by using the same equation in (2.2.2) 
and the resulting distribution is presented also in Figure 2 (dot-dashed 
curve).  We find that $\sim 16\%$ of the total Galactic bulge events with 
giant source stars will be transit events.

However, not all the source-transit events can be detected by the current 
lensing experiments.  From Figure 2 one finds that source-transit events     
tend to have Einstein time scales substantially shorter than the average value 
of the total events.  This implies that most source-transit events are caused 
by low-mass lenses.  Then the most important restriction in detecting 
source-transit events is given by the short duration of source crossing. 
For a transit event with an impact parameter $\beta$, the duration of 
source crossing is computed by
$$
t_{\rm cross} = 2(\beta_{\rm crit}^2 - \beta^2)^{1/2} t_{\rm E}.
\eqno(2.2.4)
$$
We, therefore, estimate the fraction of detectable source-transit events
by assuming that only events with $t_{\rm cross}$ longer than 2 days, which  
is the minimum duration for intensive monitoring by the followup 
observations\footnote{Since the color changes caused by the limb darkening 
effect will be small ($\sim 0.05$ mag), intensive minotoring of events with 
high precision followup observations will be essential for the detection of 
transit events.  Currently several followup observation teams are under 
operation (GMAN: Pratt et al.\ 1996; PLANET: Albrow et al.\ 1998; MPS: Rhie 
et al.\ 1999).  Since by observing giant source star events one can not only 
construct light curves with small uncertainties but also has better chance to 
detect the finite-source effect, most of these events are monitored by 
followup observations.}, can be detected.  With this defintion of detectable 
source-transit events, we compute the event rate distribution for these events 
and it is presented in Figure 3 (solid curve).  We find that source-transit 
can be detected for $\sim 2\%$ of Galctic bulge events with giant source 
stars.

\section{Color Change Caused by the Blending Effect}

The light curve of a microlensing event can also become chromatic by blending.
If one lets the blended fluxes in two different bands $B_{\nu i}$, the colors 
of a point-source event before and during the event are $(m_{\nu 2}-
m_{\nu 1})_0 = -2.5\log [(F_{\nu 2}+B_{\nu 2})/(F_{\nu 1}+B_{\nu 1})]$ and 
$m_{\nu 2}- m_{\nu 1}=-2.5\log [(A_0F_{\nu 2}+B_{\nu 2})/(A_0F_{\nu 1}+ 
B_{\nu 1})]$, respectively. Then, the color changes caused by the blending 
effect for a point-source event is computed by
$$
\Delta(m_{\nu 2}-m_{\nu 1}) = 
-2.5 \log \left[ 
\left( {A_0+f_{\nu 2} \over A_0+f_{\nu 1}}\right)
\left( {1+f_{\nu 2} \over 1+f_{\nu 1}}\right)^{-1}
\right],
\eqno(3.1)
$$
where $f_{\nu i}=B_{\nu i}/F_{\nu i}$ represent the fractions of the blended 
light in the individual wavelength bands. Then, the change in color becomes 0 
either when there is no blended light, i.e.\ $f_{\nu i} = 0$, or when the 
source star is not gravitationally amplified, i.e.\ $A_0=1.0$. While the color 
change induced by the limb darkening effect depends only on the amplification 
ratio measured in the two wavelength bands [see equation (2.5)], the color 
change caused by the blending effect depends on the amplification and the 
blended light fractions.

To see the pattern of the color changes caused by the blending effect, we 
compute the color changes expected when an example Galactic bulge event is 
affected by various fractions of blended light. For the direct comparison 
with the color changes caused by the limb-darkening effect analyzed in 
previous section, we assume the same impact parameter and the same lensed 
source star type of a K giant. For blended stars, we assume that they are 
K-type main-sequence stars, which are the most numerous type of stars in 
the Galactic bulge field. We determine the apparent magnitudes (i.e.\ $U$ 
and $I$) of the lensed and blended stars by using the data of absolute 
luminosities and colors provided by Allen (1973), Schmidt-Kaler (1982), 
and Peletier (1989) along with the adopted distance to source stars of 
$D_{os}=8$ kpc: $U=17.03$ and $I=14.05$ for the lensed giant and $U=21.67$ 
and $I=19.01$ for each of the blended main-sequence star.  Then, if a single 
main-sequence star is blended, the blended light fractions in the individual 
bands are $f_U=1.39\%$ and $f_I=1.04\%$, respectively. The individual curves 
in Figure 4 represent the color changes when the numbers of the blended 
main-sequence stars are 1, 3, 5, 10, and 20, respectively.

From Figure 1 and Figure 4, one finds that the pattern of the color changes 
caused by the blending effect is similar to the color changes caused by the
limb-darkening effect for passing events; increasing the amount of color 
change with decreasing lens-source separation. In addition, one finds that 
although the amount of color change caused by the effect of blending depends 
on the fractions of the blended light, the induced color change is equivalent 
to that caused by the limb-darkening effect.

Since the densities of stars both in the Galactic bulge and the Magellanic 
Cloud fields are very high, it is very likely that most of events detected 
towards these directions are affected by blending.  Then if the blending 
effect is not corrected for, the obtained color curve of an event will be 
seriously distorted by the color changes induced the blending effect, causing 
one to obtain wrong information both about the source star brightness profile 
and the lens proper motion. By considering both the effects of limb darkening 
and blending, the observed color change of an event is computed by
$$
\Delta(m_{\nu 2}-m_{\nu 1}) = 
-2.5 \log \left[ 
\left( {A_{\nu 2}+f_{\nu 2} \over A_{\nu 1}+f_{\nu 1}}\right)
\left( {1+f_{\nu 2} \over 1+f_{\nu 1}}\right)^{-1}
\right].
\eqno(3.2)
$$
Note that due to the limb-darkening effect the amplification is no longer 
achromatic unlike the amplification for a point-source event [cf.\ equation 
(3.1)]. In Figure 5, we present the color curve of an event distorted by the 
blending effect (solid curve), in which the flux of a lensed K-type giant 
source star is affected by the fluxes from two blended main-sequence stars. 
One finds that even for a very small fraction ($\sim 2\%$) of the blended 
light the observed color curve significantly differs from the unblended color 
curve (dotted curve) due to the color changes caused by blending (dot-dashed 
curve).

Then, how often the color curves of passing events will
be affected by the blending effect.  To see this, we compute the average 
fraction of blended light for an event with a K-type giant source star by 
adopting the luminosity function of Holtzman et al.\ (1998) constructed 
from the observations by using the {\it Hubble Space Telescope} (HST).  
From this computation, we find that the average fraction of blended light, 
which comes mostly from main-sequence stars, is considerable ($\gtrsim 5\%$) 
even for a bright giant source star.  Considering the fact that the blending 
effect on the observed light curve is significant even for $\sim 2\%$ of 
blended light, the color curves of most Galactic bulge events will be 
affected by the blending effect.

\section{Blending Correction}

In previous section, we demonstrated that since the color changes caused 
by the blending effect can be equivalent to those induced by the 
limb-darkening effect, the color curve of an event can be seriously 
distorted by the blending effect.  Therefore, to obtain useful information 
both about the lens and the source star from the color curve of the event, 
it will be essential to eliminate or correct for the blending effect. 
There have been various methods proposed for the correction of the blending 
effect. In this section, we discuss the applicabilities of these methods 
to the precise construction of color curves which are free from the blending 
effect.

First, one can correct for the blending effect by detecting the shift of a 
source star image centroid toward the blended star: {\it centroid shift method} 
(Alard, Mao, \& Guibert 1995; Alard 1996; Goldberg \& Wo\'zniak 1998; Han, 
Jeong, \& Kim 1998). Noticeable color changes induced by the limb-darkening 
effect occurs for events with giant stars. For these bright source events, the 
fraction of blended light will be small. Then, since the expected centroid 
shift caused by a faint blended star will be too small to be detected, this 
method is not appropriate for the detection and correction for the blending 
effect caused by a small fraction of blended light.

Second, the effect of blending can also be corrected for by astrometrically 
observing microlensing events with a high precision interferometer such as the 
Space Interferometry Mission (SIM): {\it astrometric method} (Han \& Kim 1999).
When a source star is gravitationally amplified, it is split into two images, 
and their center of light moves along an elliptical trajectory (astrometric 
ellipse), which can be measured from the astrometric observations by using the 
SIM, during the event (Walker 1995; H\o\hskip-1pt g, Novikov \& Polnarev1995;
Paczy\'nski 1998; Boden, Shao \& Van Buren 1998; Han \& Chang 1999). If the 
event is affected by the blending effect, on the other hand, the observed 
trajectory of the source image centroid shift will deviate from the elliptical 
one. Since the deviation of the trajectory will be significant even for a small
fraction of blended light, astrometric observations of microlensing events will
be an efficient method to correct for the blending effect. The problem of this 
method is, however, that the mission is scheduled to be launched in 2005, and 
thus it is not immediately available.

Third, with the recently developed method of the difference image analysis 
method (DIA) one can measure light variations which are free from the effect 
of blending: {\it DIA method} (Alard 1999; Alard \& Lupton 1998; Melchior 
et al.\ 1998, 1999; Alcock et al.\ 1999a, 1999b).  The DIA method detects 
and measures the variations of source star flux $\Delta F_{\nu}$ by 
subtracting an observed image from a convolved reference image.  Then, with 
the DIA method one can measure the true color of a point-source star by 
$\Delta(m_{\nu 2}-m_{\nu 1})=-2.5 \log (\Delta F_{\nu 2}/\Delta F_{\nu 1})$ 
because the ratio between the variations of source star flux measured in two 
passbands by the DIA method is identical to the flux ratio of the source star 
fluxes, i.e.\ $\Delta F_{\nu 2}/\Delta F_{\nu 1}=F_{\nu 2}/F_{\nu 1}$. 
However, the color change induced by the limb-darkening effect cannot be 
determined by the DIA method. This is because the ratio of the light variation 
for the limb-darkened extened source event is $\Delta F_{\nu 2}/\Delta 
F_{\nu 1}=(F_{\nu 2}/F_{\nu 1})[(A_{\nu 1}-1) /(A_{\nu 2}-1)]$, while the 
value required for the color measurement is 
$A_{\nu 2}F_{\nu 2}/A_{\nu 1}F_{\nu 1}$.

Fourth, one can also correct for the blending effect from the high resolution 
space observations by using the HST: {\it HST method} (Han 1997). Blending 
corrections for all events by using this method will be difficult due to a 
large number of hours of HST time.  However, if the HST observations are 
selectively conducted for relatively rare high amplification events with 
giant source stars, for which the probability to detect the limb-darkening 
effect is very high, the observations can be conducted with a reasonable 
amount of HST time. Actually, the MACHO groups already had HST observations 
of microlensing sources for a number of important events (Alcock et al.\ 
1999a).

\section{Summary}

We analyze the color changes of gravitational microlensing events caused by 
the two different mechanisms of limb darkening and blending. The findings 
from these analyses are summarized as follows.

\begin{enumerate}
\item
The color changes induced by the limb-darkening effect have dramatically 
different patterns depending on whether the lens transits the surface of the 
source star. The color curve of a source-transit event is characterized by the 
turn of the color change, while that of a passing event has no turn and 
relatively smaller amount of color change.
\item
For a source-transit event, one can measure the lens proper motion by simply 
measuring the turning time of the color curve instead of fitting the overall 
light or color curves.
\item
Among the total Galactic bulge events with giant source stars, $\sim 16\%$ of 
them are expected to be source-transit events and $\sim 2\%$ of events will 
have source-srossing intervals long enough to be detectable by the current 
followup observations.
\vskip 1cm
\item
Even for a very small fraction of blended light, the color change caused by 
the blending effect can become equivalent to those induced by the 
limb-darkening effect, causing serious distortion in the observed color curve 
of the event. Therefore, to obtain useful information both about the lens 
and source star, it will be essential to correct for the blending effect.
\item
Among the methods proposed to correct for the blending effect, the best 
solution for the precise construction of the color curve is provided by the 
HST observations of events.
\end{enumerate}

\acknowledgements
This work was supported by the grant (1999-2-113-001-5) of the Korea Science 
\& Engineering Foundation.

\clearpage
\postscript{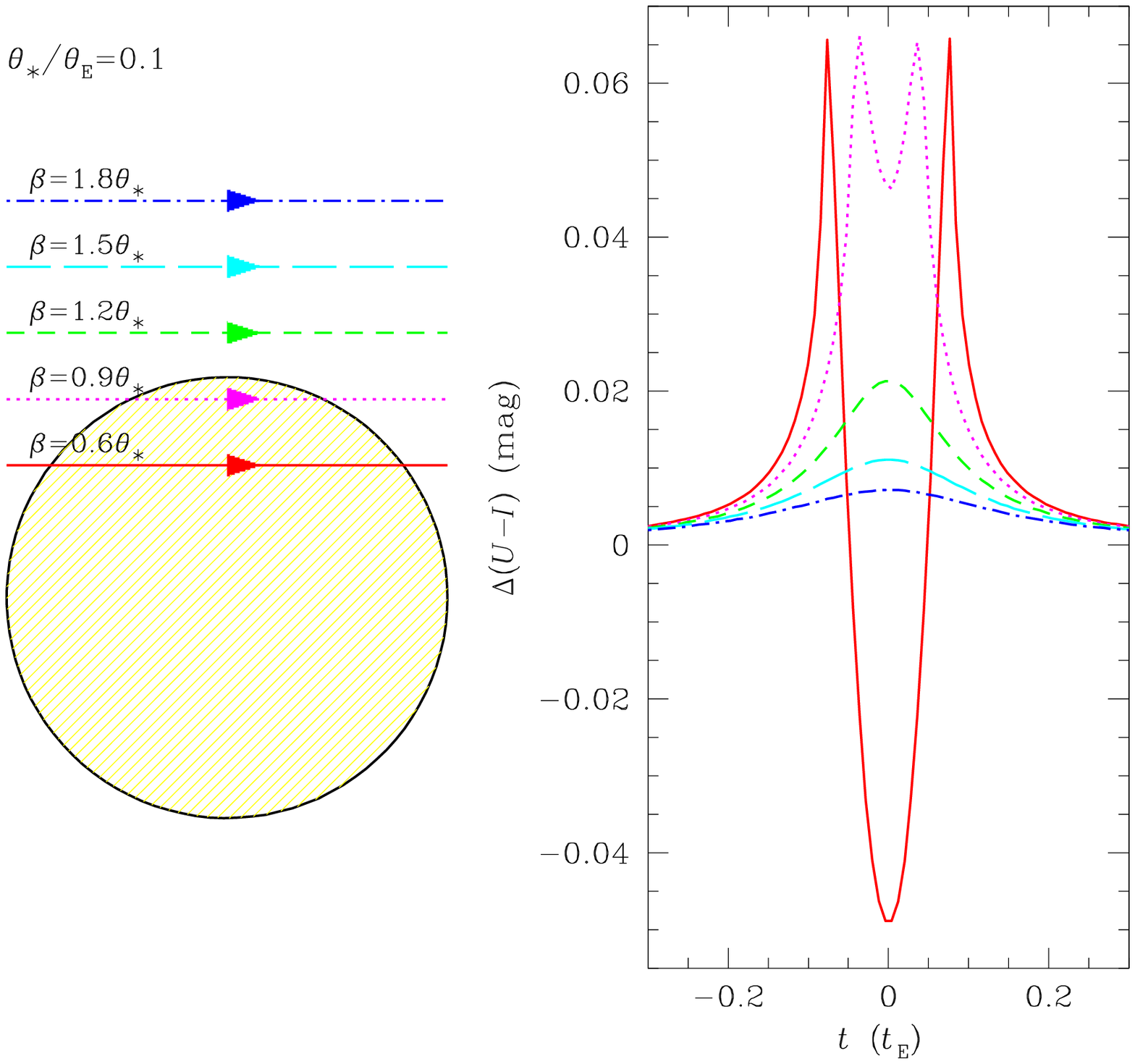}{1.0}
\noindent
{\footnotesize {\bf Figure 1:}\ 
The color changes of microlensing events with a limb-darkened extended source 
star. The source star has an angular radius of $\theta_\ast=0.1\theta_{\rm E}$.
On the left side, the trajectories of the lens with respect to the source star
(the shaded circle) are marked by straight lines with various line types. In 
the right panel, the resulting color curves $\Delta(U-I)$ for the individual 
events are drawn by the same line types as those of the corresponding lens 
trajectories in the left panel. For the brightness profiles of the source star
in the two wavelength bands, see the text.
}\clearpage

\postscript{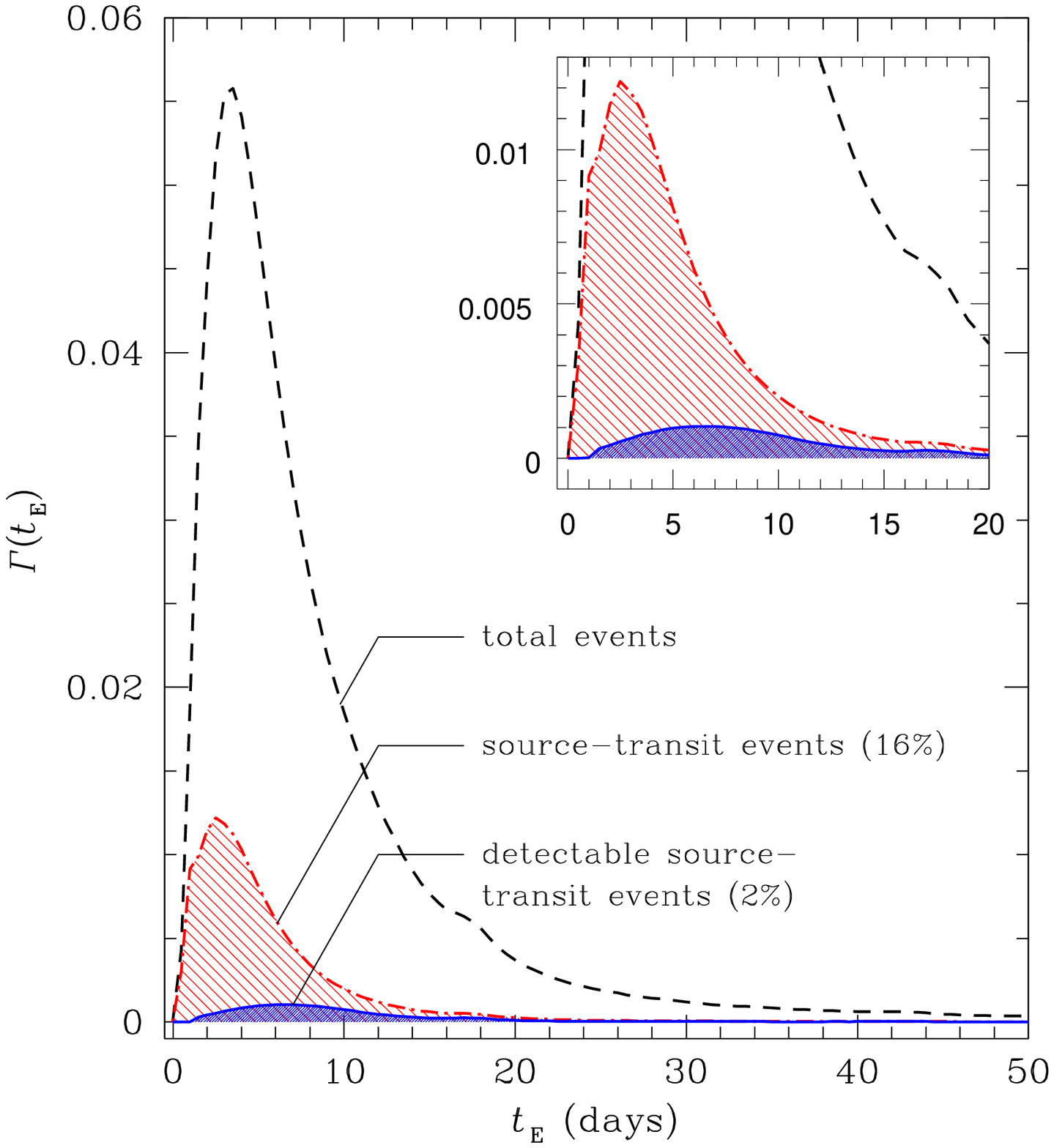}{1.0}
\noindent
{\footnotesize {\bf Figure 2:}\ 
The event rate distributions of Galactic bulge events as a function of 
Einstein time scale.  While the dashed curve represents the distribution 
for the total events with giant source stars, dot-dashed curve is only for 
source-transit events.  The solid curve represents the distribution for 
detectable source-transit events.  The total event rate is arbitrarily 
normalized, but the individual distributions are relatively scaled.  To 
better show the distribution for transit events, the region with short 
$t_{\rm E}$ is expanded and presented in a separate box.
}\clearpage

\postscript{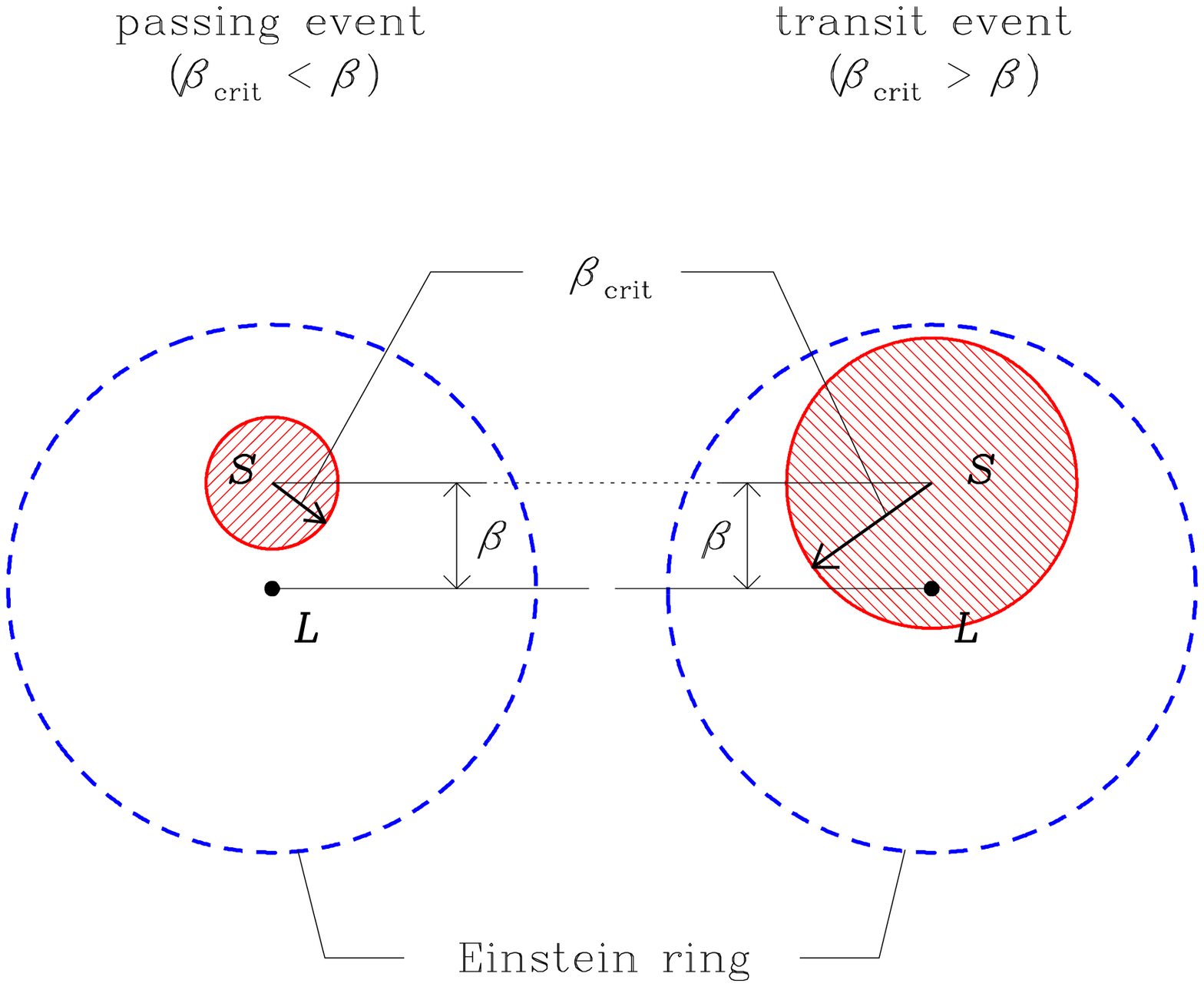}{1.0}
\noindent
{\footnotesize {\bf Figure 3:}\ 
The lens system geometries for passing and transit events.  The dotted 
circle on each side represents the Einstein ring and the lens ($L$) is 
located at the center of the Einstein ring.  The surface of the source 
star ($S$) is represented by a shaded circle.  The value $\beta_{\rm crit}$ 
represents the source star radius normalized by $\theta_{\rm E}$.  One finds 
that to become a source-transit event $\beta_{\rm crit}$ should be greater 
than the impact parameter ($\beta$) of the event.
}\clearpage

\postscript{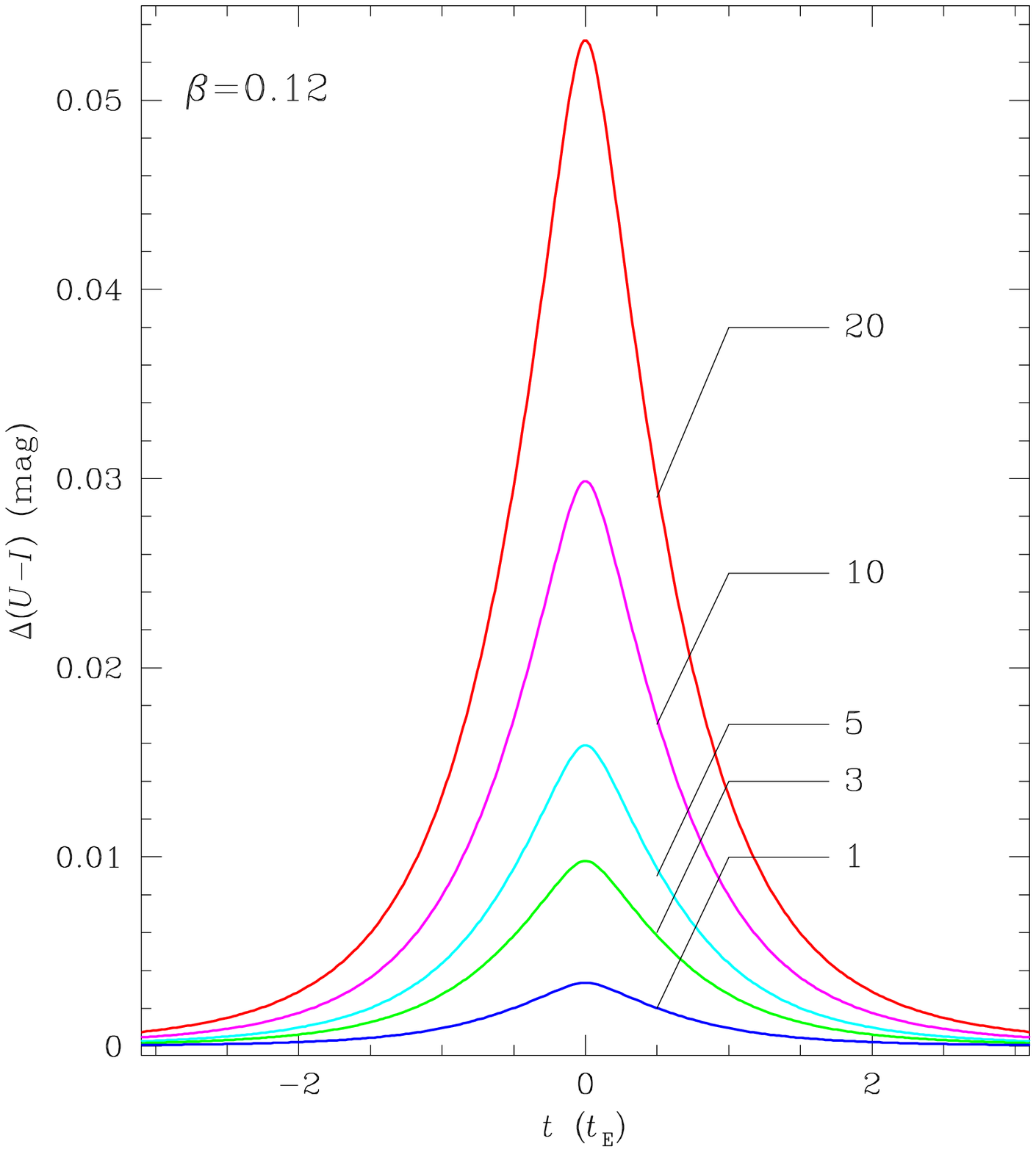}{1.0}
\noindent
{\footnotesize {\bf Figure 4:}\ 
The color changes of microlensing events affected by various fractions of 
blended light. The lensed star is a K-type giant with $U=17.03$ and $I=14.05$, 
while the individual blended stars are assumed to be main-sequence stars with 
$U=21.67$ and $I=19.01$. The individual curves represent the color changes when
the numbers of main-sequence stars contributing to blended light are 1, 3, 5, 
10, and 20, respectively. The blended light fractions by each blended star in 
the individual wavelength bands are $F_U=1.39\%$ and $F_I=1.04\%$. The marked 
number for each curve represents the number of blended main-sequence stars.
}\clearpage

\postscript{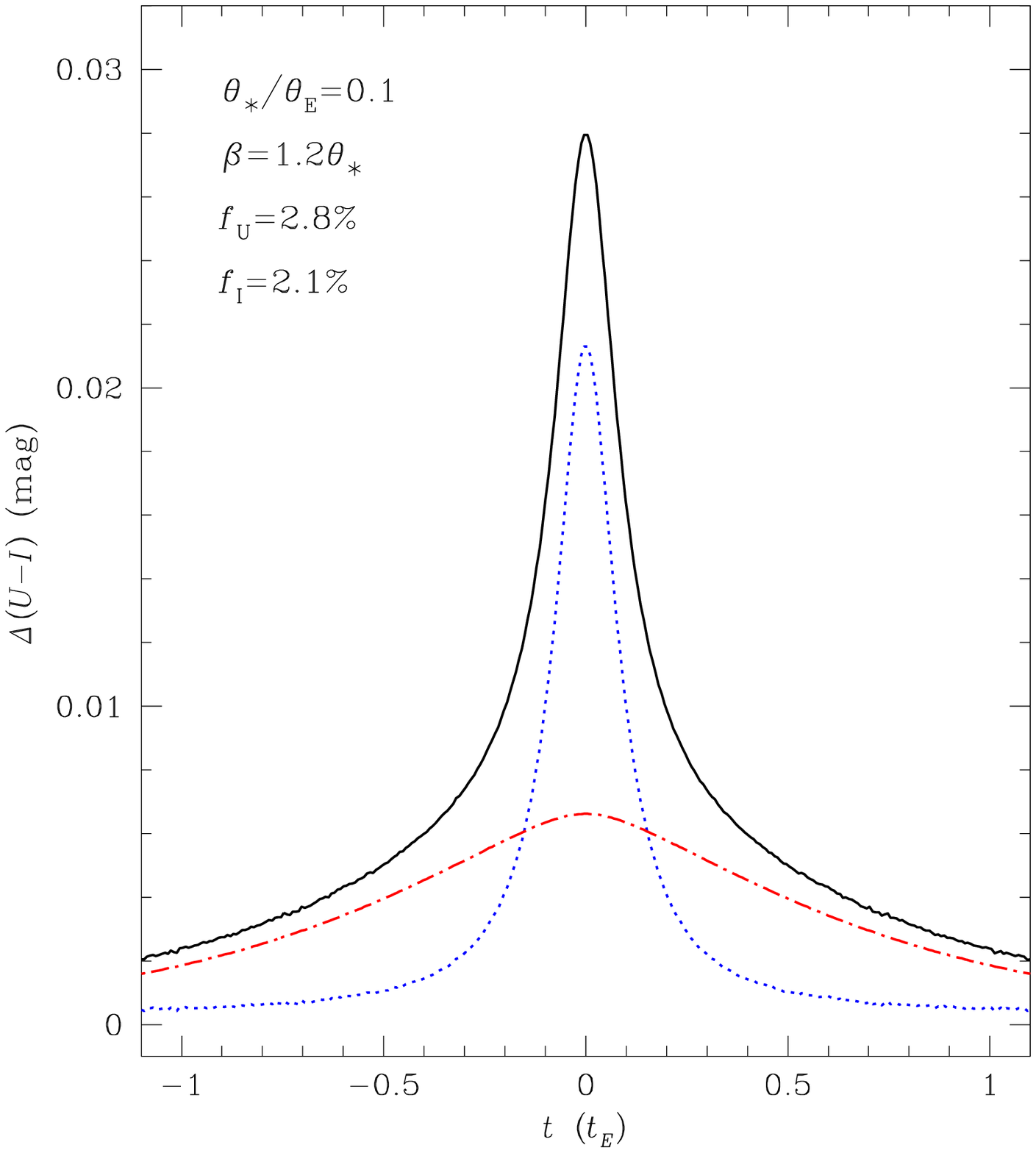}{1.0}
\noindent
{\footnotesize {\bf Figure 5:}\ 
The color curve of an event affected by blending (solid curve). The dotted 
curve represents the color curve expected when the event is not affected by 
the blending effect, while the color changes caused purely by the blending 
effect is marked by a dot-dashed curve. The source star radius 
($\theta_\ast/\theta_{\rm E}$), the impact parameter of the event ($\beta$), 
and the fractions of blended light in the individual bands ($f_U$ and $f_I$) 
are also marked.
}\clearpage

\end{document}